\documentclass{aa}  
\usepackage{graphicx}
\usepackage{natbib}
\usepackage{aalongtable}
\usepackage{subfigure}

\bibpunct{(}{)}{;}{a}{}{,}   
\usepackage{txfonts}

\begin{document}
\title{Dust and gas emission in the prototypical hot
  core G29.96$-$0.02 at sub-arcsecond resolution}


   \author{H.~Beuther\inst{1}, Q. Zhang\inst{2}, E.A. Bergin\inst{3}, T.K. Sridharan\inst{2}, T.R. Hunter\inst{4}, S. Leurini\inst{5}}


   \offprints{H.~Beuther}

   \institute{Max-Planck-Institute for Astronomy, K\"onigstuhl 17, 
              69117 Heidelberg, Germany\\
              \email{beuther@mpia.de}
         \and
              Harvard-Smithsonian Center for Astrophysics, 60 Garden Street,
              Cambridge, MA 02138, USA\\
             \email{name@cfa.harvard.edu}
         \and 
              University of Michigan,  Dept. of Astronomy,
              Ann Arbor, MI 48109-1090\\
              \email{ebergin@umich.edu}
         \and 
              NRAO, 520 Edgemont Rd,
              Charlottesville, VA 22903\\
              \email{thunter@nrao.edu}
         \and
              European Southern Observatory, Karl-Schwarzschild-Str. 2, 
              85748 Garching, Germany\\
              \email{sleurini@eso.org}
             }
\authorrunning{Beuther et al.}
\titlerunning{SMA observations of G29.96$-$0.02}

   \date{}

  \abstract
  {Hot molecular cores are an early manifestation of massive star
    formation where the molecular gas is heated to temperatures
    $>100$\,K undergoing a complex chemistry.}
  {One wants to better understand the physical and
    chemical processes in this early evolutionary stage.}
  {We selected the prototypical hot molecular core G29.96$-$0.02 being
    located at the head of the associated ultracompact H{\sc
      ii} region. The 862\,$\mu$m submm continuum and spectral line
    data were obtained with the Submillimeter Array (SMA) at
    sub-arcsecond spatial resolution.}
  {The SMA resolved the hot molecular core into six submm continuum
    sources with the finest spatial resolution of $0.36''\times
    0.25''$ ($\sim$1800\,AU) achieved so far. Four of them located
    within 7800\,(AU)$^2$ comprise a proto-Trapezium system with
    estimated protostellar densities of $1.4\times 10^5$
    protostars/pc$^3$. The plethora of $\sim$ 80 spectral lines allows
    us to study the molecular outflow(s), the core kinematics, the
    temperature structure of the region as well as chemical effects.
    The derived hot core temperatures are of the order 300\,K. We find
    interesting chemical spatial differentiations, e.g., C$^{34}$S is
    deficient toward the hot core and is enhanced at the UCH{\sc ii}/
    hot core interface, which may be explained by temperature
    sensitive desorption from grains and following gas phase
    chemistry. The SiO(8--7) emission outlines likely two molecular
    outflows emanating from this hot core region.  Emission from most
    other molecules peaks centrally on the hot core and is not
    dominated by any individual submm peak.  Potential reasons for
    that are discussed. A few spectral lines that are associated with
    the main submm continuum source, show a velocity gradient
    perpendicular to the large-scale outflow.  Since this velocity
    structure comprises three of the central protostellar sources,
    this is not a Keplerian disk. While the data are consistent with a
    gas core that may rotate and/or collapse, we cannot exclude the
    outflow(s) and/or nearby expanding UCH{\sc ii} region as possible
    alternative causes of this velocity pattern.}
  {}

   \keywords{ stars: formation -- ISM: jets and outflows -- ISM:
     molecules -- stars: early-type -- stars: individual (G29.96$-$0.02) --
     (stars:) binaries (including multiple): close}

   \maketitle
%

\section{Introduction}

Hot molecular cores represent an early evolutionary stage in massive
star formation prior to the formation of an ultracompact H{\sc ii}
region (UCH{\sc ii}). Single-dish line surveys toward hot cores have
revealed high abundances of many molecular species and temperatures
usually exceeding 100\,K (e.g.,
\citealt{schilke1997a,hatchell1998b,mccutcheon2000}).  Unfortunately,
most hot cores are relatively far away (a few kpc, Orion-KL being an
important exception), and high-spatial resolution studies are
important to disentangle the various components in the region, to
resolve potential multiple heating sources, and to search for chemical
variations throughout the regions. Here we present sub-arcsecond
resolution submm spectral line and dust continuum observations of the
hot core G29.96$-$0.02, characterizing the physical and chemical
properties of this prototypical region.

The hot core/UCH{\sc ii} region G29.96$-$0.02 is a well studied source
comprising a cometary UCH{\sc ii} region and approximately $2.6''$ to
the west a hot molecular core
(\citealt{wc1989b,cesaroni1994,cesaroni1998}). G29.96$-$0.02 is at a
distance of $\sim$6\,kpc \citep{pratap1999}, the bolometric luminosity
measured with IRAS is very high with $L\sim 1.4\times
10^6$\,L$_{\odot}$ \citep{cesaroni1994}. Since the region harbors at
least two massive (proto)stars (within the UCH{\sc ii} region and the
hot core) this luminosity must be distributed over various sources.
Based on cm continuum free-free emission, \citet{cesaroni1994}
calculate a luminosity for the UCH{\sc ii} region of $L_{\rm{cm}}\sim
4.4\times 10^5$\,L$_{\odot}$. Furthermore, they try to estimate the
luminosity of the hot core via a first order black-body approximation
and get a value of $L_{\rm{bb}}\sim 1.2\times 10^5$\,L$_{\odot}$.
Later, \citet{olmi2003} derive a similar estimate ($\sim 9\times
10^4$\,L$_{\odot}$) via integrating a much better determined SED. The
exciting source of the UCH{\sc ii} region has been identified in the
near-infrared as an O5-O8 star \citep{watson1997}. Furthermore,
\citet{pratap1999} identified two additional sources toward the rim of
the UCH{\sc ii} region and an enhanced density of reddened sources
indicative of an embedded cluster.

A line survey toward a number of UCH{\sc ii} regions reveals that
G29.96$-$0.02 is a strong molecular line emitter in nearly all
observed species \citep{hatchell1998b}. High-angular resolution
studies show that many species (e.g., NH$_3$, CH$_3$CN, HNCO,
HCOOCH$_3$) peak toward the main H$_2$O maser cluster $\sim 2.6''$
west of the UCH{\sc ii} region (e.g,
\citealt{hofner1996,cesaroni1998,olmi2003}), whereas CH$_3$OH peaks
$\sim 4''$ further south-west associated with another isolated H$_2$O
maser feature \citep{pratap1999}. \citet{hoffman2003} detected one of
the relatively rare H$_2$CO masers toward the hot core position. These
masers are proposed to trace the warm molecular gas in the vicinity of
young forming massive stars \citep{araya2006}. The signature of a
CH$_3$OH peak offset from the other molecular lines is reminiscent of
Orion-KL (e.g., \citealt{wright1996,beuther2005a}).  Temperature
estimates toward the hot core based on high-density tracers vary
between 80 and 150\,K (e.g.,
\citealt{cesaroni1994,hatchell1998b,pratap1999,olmi2003}).

While \citet{gibb2004} detect a molecular outflow in H$_2$S emanating
from the hot core in approximately the south-east north-west
direction, \citet{cesaroni1998} and \citet{olmi2003} detect a velocity
gradient in the east-west direction in the high-density tracers
NH$_3$(4,4) and CH$_3$CN, consistent with a rotating disk around an
embedded protostar. However, \citet{maxia2001} also report that their
rather low-resolution $5.9''\times3.7''$ ($\approx$\,0.15\,pc)
SiO(2--1) data are consistent with the disk scenario as well. This is
a bit puzzling since SiO is usually found to trace shocked gas in
outflows and not more quiescent gas in disks. Inspecting their SiO
image again (Fig.~6 in \citealt{maxia2001}), this interpretation is
not unambiguous, the data also appear to be consistent with the
outflow observed in H$_2$S \citep{gibb2004}. It is possible that the
spatial resolution of their SiO(2--1) observations is not sufficient
to really disentangle the outflow in this distant region.

\citet{olmi2003} compiled the SED from cm to mid-infrared wavelengths.
While the 3\,mm data are still strongly dominated by the free-free
emission \citep{olmi2003}, at 1\,mm the hot core becomes clearly
distinguished from the adjacent UCH{\sc ii} region
\citep{wyrowski2002}.  G29.96$-$0.02 is one of the few hot cores which
is detected at mid-infrared wavelengths \citep{debuizer2002}.
Interestingly, the mid-infrared peak is $\sim 0.5''$ ($\sim$3000\,AU)
offset from the NH$_3$(4,4) hot core position. While \citet{gibb2004}
speculate that the mid-infrared peak might arise from the scattered
light only, \citet{debuizer2002} suggest that it could trace a second
massive source within the same core. This hypothesis can be tested via
very-high-angular-resolution submm continuum studies.

\section{Observations}
\label{obs}

We have observed the hot core G29.96$-$0.02 with the Submillimeter
Array (SMA\footnote{The Submillimeter Array is a joint project between
  the Smithsonian Astrophysical Observatory and the Academia Sinica
  Institute of Astronomy and Astrophysics, and is funded by the
  Smithsonian Institution and the Academia Sinica.}, \citealt{ho2004})
during four nights between May and November 2005. We used all
available array configurations (compact, extended, very extended, for
details see Table \ref{config}) with unprojected baselines between 16
and 500\,m, resulting at $862\,\mu$m in a projected baseline range
from 16.5 to 591\,k$\lambda$. The chosen phase center was the peak
position of the associated UCH{\sc ii} region R.A. [J2000.0]:
$18^h46^m03.^s99$ and Decl. [J2000.0] $-02^{\circ}39'21.''47$. The
velocity of rest is $v_{lsr}\sim +98$\,km\,s$^{-1}$
\citep{churchwell1990}.

\begin{table}[htb]
\caption{Observing parameters}
\begin{tabular}{lrrrr}
\hline \hline
Date & Config. & \# ant. & Source loop & $\tau(225\rm{GHz})$ \\
     &         &         & [hours] \\
\hline
28.May05 & very ext. & 6 & 7.0 & 0.13-0.16 \\
18.Jul.05& comp.     & 7 & 7.5 & 0.06-0.09 \\
4.Sep.05 & ext.      & 6 & 4.5 & 0.06-0.08 \\
5.Nov.05 & very ext. & 7 & 3.0 & 0.06\\
\hline \hline
\end{tabular}
\label{config}
\end{table}

For bandpass calibration we used Ganymede in the compact configuration
and 3C279 and 3C454.3 in the extended and very extended configuration.
The flux scale was derived in the compact configuration again from
observations of Ganymede. For two datasets of the more extended
configurations, we used 3C454.3 for the relative scaling between the
various baselines and then scaled that absolutely via observations of
Uranus.  For the fourth dataset we did the flux calibration using
3C279 only.  The flux accuracy is estimated to be accurate within
20\%. Phase and amplitude calibration was done via frequent
observations of the quasars 1743-038 and 1751+096, about
15.5$^{\circ}$ and 18.3$^{\circ}$ from the phase center of
G29.96$-$0.02. The zenith opacity $\tau(\rm{348GHz})$, measured with
the NRAO tipping radiometer located at the Caltech Submillimeter
Observatory, varied during the different observation nights between
$\sim$0.15 and $\sim$0.4 (scaled from the 225\,GHz measurement). The
receiver operated in a double-sideband mode with an IF band of
4-6\,GHz so that the upper and lower sideband were separated by
10\,GHz. The central frequencies of the upper and lower sideband were
348.2 and 338.2\,GHz, respectively. The correlator had a bandwidth of
2\,GHz and the channel spacing was 0.8125\,MHz.  Measured
double-sideband system temperatures corrected to the top of the
atmosphere were between 110 and 800\,K, depending on the zenith
opacity and the elevation of the source. Our sensitivity was
dynamic-range limited by the side-lobes of the strongest emission
peaks and thus varied between the line maps of different molecules and
molecular transitions. This limitation was mainly due to the
incomplete sampling of short uv-spacings and the presence of extended
structures.  The $1\sigma$ rms for the velocity-integrated molecular
line maps (the velocity ranges for the integrations were chosen for
each line separately depending on the line-widths and intensities)
ranged between 36 and 76\,mJy. The average synthesized beam of the
spectral line maps was $0.65'' \times 0.48''$ (P.A. $-83^{\circ}$).
The 862\,$\mu$m submm continuum image was created by averaging the
apparently line-free parts of the upper sideband. The $1\sigma$ rms of
the submm continuum image was $\sim 21$\,mJy/beam, and the achieved
synthesized beam was $0.36''\times 0.25''$ (P.A.  $18^{\circ}$), the
smallest beam obtained so far with the SMA. The different synthesized beams
between line and continuum maps are due to different applied
weightings in the imaging process (``robust'' parameters set in MIRIAD
to 0 and -2, respectively) because there was insufficient
signal-to-noise in the line data obtained in the very extended
configuration. The initial flagging and calibration was done with the
IDL superset MIR originally developed for the Owens Valley Radio
Observatory \citep{scoville1993} and adapted for the SMA\footnote{The
  MIR cookbook by Charlie Qi can be found at
  http://cfa-www.harvard.edu/$\sim$cqi/mircook.html.}. The imaging and
data analysis were conducted in MIRIAD \citep{sault1995}.

\section{Results}

\subsection{Submillimeter continuum emission}
\label{continuumtext}

\begin{figure*}[htb]
\includegraphics[angle=-90,width=18cm]{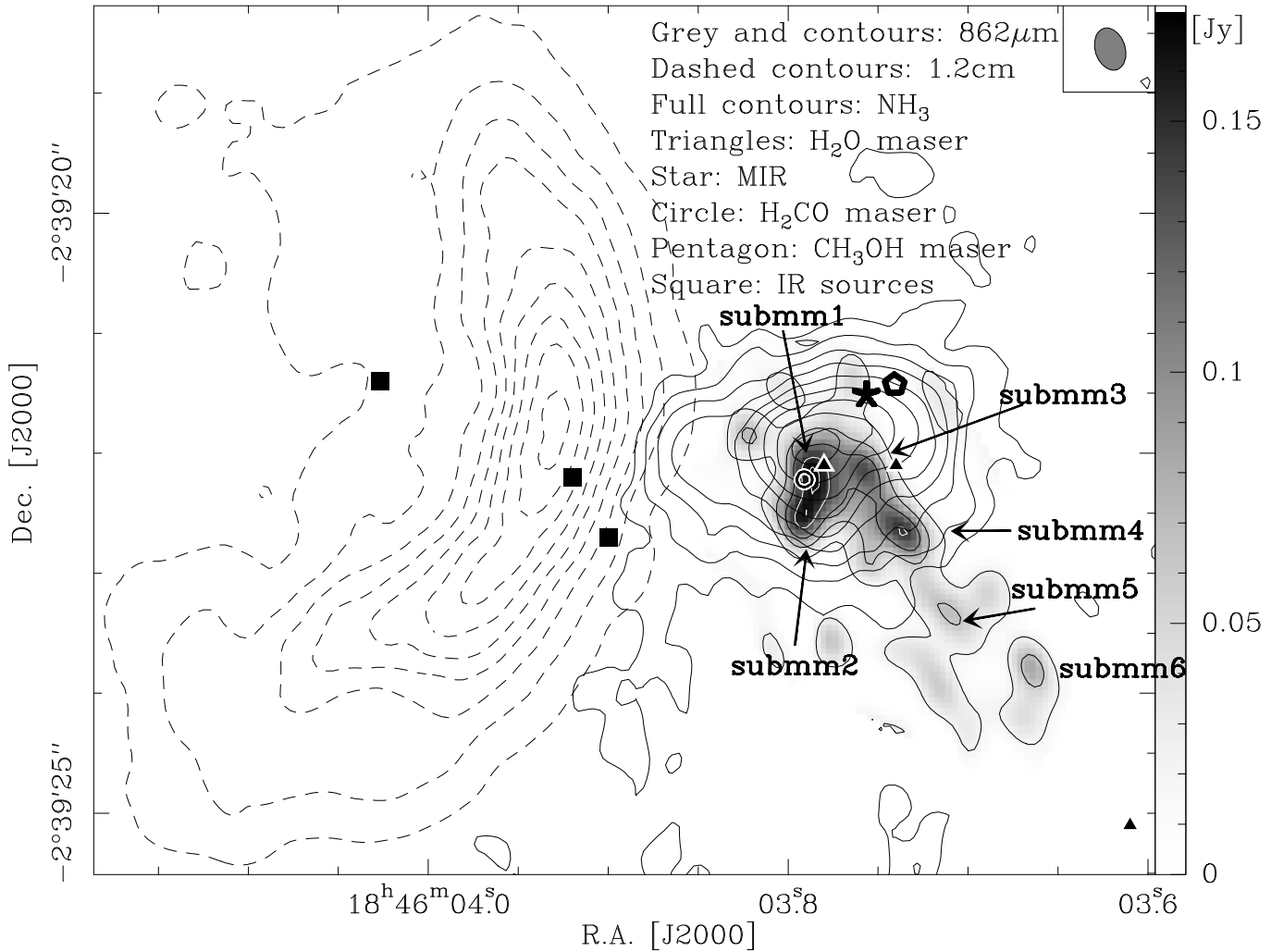}
\caption{The hot core UCH{\sc ii} region G29.96$-$0.02. The grey-scale
  with contours shows the submm continuum emission with a spatial
  resolution of $0.36''\times 0.25''$. The contour levels start at the
  1$\sigma$ level of 21\,mJy\,beam$^{-1}$ and continue at 63,
  105\,mJy\,beam$^{-1}$ (black contours) to 147, 168\,mJy\,beam$^{-1}$
  (white contours). The dashed contours outline the cm continuum
  emission from the UCH{\sc ii} region and the thick contours show the
  NH$_3$ emission \citep{cesaroni1994}.  The contouring is done from
  15 to 95\% (step 10\%) of the peak emission of each image,
  respectively ($S_{\rm{peak}}(1.2\rm{cm})=109$mJy/beam,
  $S_{\rm{peak}}(\rm{NH_3})=15$mJy/beam). Triangles, circles and
  pentagons show the H$_2$O \citep{hofner1996}, H$_2$CO
  \citep{hoffman2003} and class {\sc ii} CH$_3$OH \citep{minier2001}
  maser positions. The star marks the peak of the MIR emission
  \citep{debuizer2002}, which is not a point source but has a similar
  size as the NH$_3$ emission. The squares mark the infrared sources
  by \citet{pratap1999}.}
\label{continuum}
\end{figure*}

Figure \ref{continuum} presents the 862\,$\mu$m continuum emission
extracted from the line-free parts of the upper sideband spectrum
($\sim$1.8\,GHz in total used) shown in Figure \ref{spectra}. The very
high spatial resolution of $0.36''\times 0.25''$ corresponds to a
linear resolution of $\sim 1800$\,AU at the given distance of
$\sim$6\,kpc. The submm continuum emission peaks approximately $2''$
west of the UCH{\sc ii} region and is associated with the molecular
line emission known from previous observations. We do not detect any
submm continuum emission toward the UCH{\sc ii} region itself. At the
given spatial resolution, for the first time multiplicity within the
G29.96$-$0.02 hot core is resolved and we identify 6 submm continuum
emission peaks (submm1 to submm6) above the 3$\sigma$ level of
63\,mJy\,beam$^{-1}$ (Fig.~\ref{continuum}). We consider submm1 and
submm2 to be separate sources instead of a dust ridge because we count
compact spherical or elliptical sources and their emission peaks are
separated by about one synthesized beam. The four strongest submm
peaks, that are all $>6\sigma$ detections, are located within a region
of $(1.3'')^2$ (7800\,AU) in diameter. The submm peak submm1 is
associated with H$_2$O and H$_2$CO maser emission
\citep{hofner1996,hoffman2003}, and we consider this to be probably
the most luminous sub-source. The other H$_2$O maser peaks are offset
from the submm continuum emission. The mid-infrared source detected by
\citet{debuizer2002} is offset $>1''$ from the submm emission. This
may either be due to uncertainties in the MIR astrometry or the MIR
emission may trace another young source in the region. It should be
noted that the class {\sc ii} CH$_3$OH masers detected by
\citet{minier2001} peak close to the MIR source as well, which
indicates that the MIR offset from the hot core may well be real.

\begin{table*}[htb]
\caption{Submm continuum source parameters}
\begin{tabular}{lrrrrrr}
\hline \hline
Source & R.A. & Dec. & $S_{\rm{peak}}$ & $S_{\rm{int}}$ & $M$ & $N$ \\
       & [J2000] & [J2000] & [$\frac{\rm{mJy}}{\rm{beam}}$] & [mJy] & [M$_{\odot}]$ & [10$^{24}$cm$^{-3}$]\\
\hline
submm1 & 18:46:03.786 & -02:39:22.19 & 173 & 288 & 11.5 & 5.7 \\
submm2 & 18:46:03.789 & -02:39:22.48 & 168 & 237 & 9.5 & 5.5 \\
submm3 & 18:46:03.758 & -02:39:22.16 & 138 & 178 & 7.1 & 4.5 \\
submm4 & 18:46:03.736 & -02:39:22.65 & 151 & 249 & 9.9 & 5.0 \\
submm5 & 18:46:03.710 & -02:39:23.33 &  68 & 106 & 4.2 & 2.2 \\
submm6 & 18:46:03.665 & -02:39:23.80 &  84 &  85 & 3.4 & 2.8 \\
\hline \hline
\end{tabular}
~\\
\footnotesize{The Table shows the peak intensities $S_{\rm{peak}}$, the integrated intensities $S_{\rm{int}}$, the derived gas masses $M$ as well as the H$_2$ column densities $N$.}
\label{submmcont}
\end{table*}

Table \ref{submmcont} lists the absolute source positions, their
862\,$\mu$m peak intensities and the integrated flux densities
approximately associated with each of the sub-sources. Calculating the
brightness temperature $T_b$ of the corresponding Planck-function for,
e.g., submm1, we get $T_b(\rm{Peak1})\sim 27$\,K. Assuming hot core
dust temperatures of $\sim$ 100\,K, the usual assumption of optically
thin dust emission is not really valid anymore, and one gets an
approximate beam-averaged optical depth $\tau$ of the dust emission of
$\sim$0.3. To calculate the dust and gas masses, we can follow the
mass determination outlined in \citet{hildebrand1983} and
\citet{beuther2002a,beuther2002erratum}, which assumes optically thin
emission, and correct that for the increased dust opacity. Assuming
constant emission along the line of sight, the opacity correction
factor $C$ is
$$ C = \frac{\tau}{1-e^{-\tau}}. $$
With $\tau\sim 0.3$, we get a correction factor $C\sim 1.16$ still
comparably small. Assuming a dust opacity index $\beta = 1.5$, the
dust opacity per unit dust mass is $\kappa(862\mu\rm{m}) \sim
1.5$\,cm$^2$g$^{-1}$ (with the reference value $\kappa(250\mu\rm{m})
\sim 9.4$\,cm$^2$g$^{-1}$, see \citealt{hildebrand1983}), and we
assume a gas-to-dust ratio of 100. Given the uncertainties in $\beta$
and $T$, we estimate the masses to be accurate within a factor 4.
Table \ref{submmcont} gives the derived masses and beam-averaged
column densities. Each sub-peak has a mass of a few $M_{\odot}$, and
the main submm1 exhibits approximately 10\,$M_{\odot}$ of compact,
warm gas and dust emission.  The integrated 862\,$\mu$m continuum flux
density of the central region comprising the four main submm continuum
sources amounts to 1.16\,Jy.  At an average dust temperature of
100\,K, this corresponds to a central core mass of 39.9\,M$_{\odot}$.
In comparison to these flux density measurements, \citet{thompson2006}
observed with SCUBA 850\,$\mu$m peak and integrated flux densities of
$\sim 11.5\pm 1.2$\,Jy/($14''$beam) and $\sim$19.2\,Jy, respectively.
The ratio between peak and integrated JCMT fluxes already indicates
non-compact emission even on that scales.  Furthermore, subtracting a
typical line contamination of the continuum emission in hot cores of
the order 25\% (e.g., NGC6334I, Hunter et al.~in prep.), the total
850\,$\mu$m single-dish continuum flux density should amount to
$\sim$8.6\,Jy.  Compared with the integrated flux density in the SMA
data of $\sim$1.74\,Jy, this indicates that approximately 80\% of the
single-dish emission is filtered out by the missing short spacings in
the interferometer data. The dust and gas in the central region have
higher temperatures than the components filtered out on larger spatial
scales, and since the dust and gas mass is inversely proportionally
related to the temperature by $M_{\rm{H_2}} \propto (e^{h\nu/kT}-1)$
(e.g., \citealt{beuther2002a}), a greater proportion of the mass
($>80\%$) is filtered out in the SMA data.  However, the SMA image
reveals the most compact hot gas and dust cores at the center of the
evolving massive star-forming region. The shortest baseline of the SMA
observations of $\sim$16.5\,k$\lambda$ correspond to scales $>12''$
which hence have to be filtered out entirely. However, even smaller
scales are missing because the uv-spacings corresponding to scales
$\geq 5''$ are still relatively poorly sampled and the image presented
in Figure \ref{continuum} is only sensitive to spatial scales of the
order a few arcseconds. The submm peaks detected by the SMA are much
stronger than what would have been expected if the single-dish flux
($\sim$8.6\,Jy) were uniformly distributed over the SCUBA primary beam
of $14''$, even ignoring any spatial filtering and missing flux
effects (This would result in $\sim 4$\,mJy per synthesized SMA
beam.).  This shows that the emission measured on the small spatial
scales sampled by the SMA represents the compact core emission much
better than expected. However, it does not imply that the gas masses
measured by the SMA are the only gas reservoir the embedded protostars
have for their ongoing accretion; they may also gain mass from the
large-scale gas envelope that is filtered out by our observations (see
also the competitive accretion scenario, e.g., \citealt{bonnell2004}).
The derived beam-averaged H$_2$ column densities are of the order a
few times $10^{24}$\,cm$^{-2}$, corresponding to visual extinctions
$A_v$ of a few 1000 ($A_v=N_{\rm{H}}/0.94\times 10^{21}$,
\citealt{frerking1982}).
 
\subsection{Spectral line emission}

Figure \ref{spectra} presents spectra extracted toward the main submm
submm1 with an angular resolution of $0.64''\times 0.47''$ compared to
the submm continuum map (see \S\ref{obs}).  More than 80 spectral
lines from 18 molecular species, isotopologues or vibrationally
excited species have been identified with a minor fraction of
$\sim$5\% of unidentified lines (UL) (Tables \ref{linelist} \&
\ref{species}). The range of upper level excitation temperatures for
the many lines varies between approximately 40 and 750\,K (Table
\ref{linelist}). Therefore, with one set of observations we are able
to trace various gaseous temperature components from the relatively
colder gas surrounding the hot core region to the densest and warmest
gas best observed in some of the vibrationally excited lines.

\begin{table}[htb]
  \caption{Peak intensities, rms and velocity ranges for images in Figs.~\ref{lineimages} \& \ref{lineimages2}.}
\begin{tabular}{lrrr}
\hline \hline
Line & $S_{\rm{peak}}$ & rms & $\Delta v$ \\
    & mJy/beam     & mJy/beam & km/s \\
\hline 
862$\mu$m cont., low res. & 422 & 17 \\
CH$_3$OH$(7_{3,5}-6_{2,4})$ & 878 & 64 & [90,104] \\
$^{13}$CH$_3$OH$(13_{7,7}-12_{7,6})$ & 752 & 51 & [95,101] \\
CH$_3$OH$(7_{4,3}-6_{4,3}), v_t=1$ & 1419 & 69 & [91,105]  \\
CH$_3$OCH$_3(7_{4,3}-6_{3,4})$ & 669 & 46 & [94,104] \\
C$_2$H$_5$OH$(15_{7,9}-15_{6,10})$ & 586 & 51 & [95,100] \\
SiO$(8-7)$ & 391 & 36 & [75,105] \\
C$^{34}$S$(7-6)$ & 592 & 62 & [92,104] \\
H$_2$CS$(10_{1,0}-9_{1,9})$ & 933 & 69 & [92,100] \\
$^{34}$SO$(8_8-7_7)$ & 827 & 57 & [95,103] \\
SO$_2(14_{4,14}-18_{3,15})$ & 544 & 53 & [94,100] \\
HCOOCH$_3(27_{5,22}-26_{5,21})$ & 491 & 70 & [96,100] \\
CH$_3$CN$(19_8-18_8)$ & 788 & 71 & [94,100] \\
CH$_3$CH$_2$CN$(38_{3,36}-37_{3,35})$ & 791 & 56 & [94,102] \\ 
CH$_3$CHCN$(36_{2,34}-35_{2,32})$ & 655 & 68 & [96,100] \\
HC$_3$N$(37-36), v_7=1$ & 622 & 55 & [94,102] \\ 
HC$_3$N$(37-36), v_7=2$ & 416 & 57 & [94,100] \\
HN$^{13}$C$(4-3)$ & 1149 & 76 & [94,100] \\ 
\hline \hline
\end{tabular}
\label{rms}
\end{table}

\begin{figure*}[htb]
\begin{center}
\includegraphics[angle=-90,width=18cm]{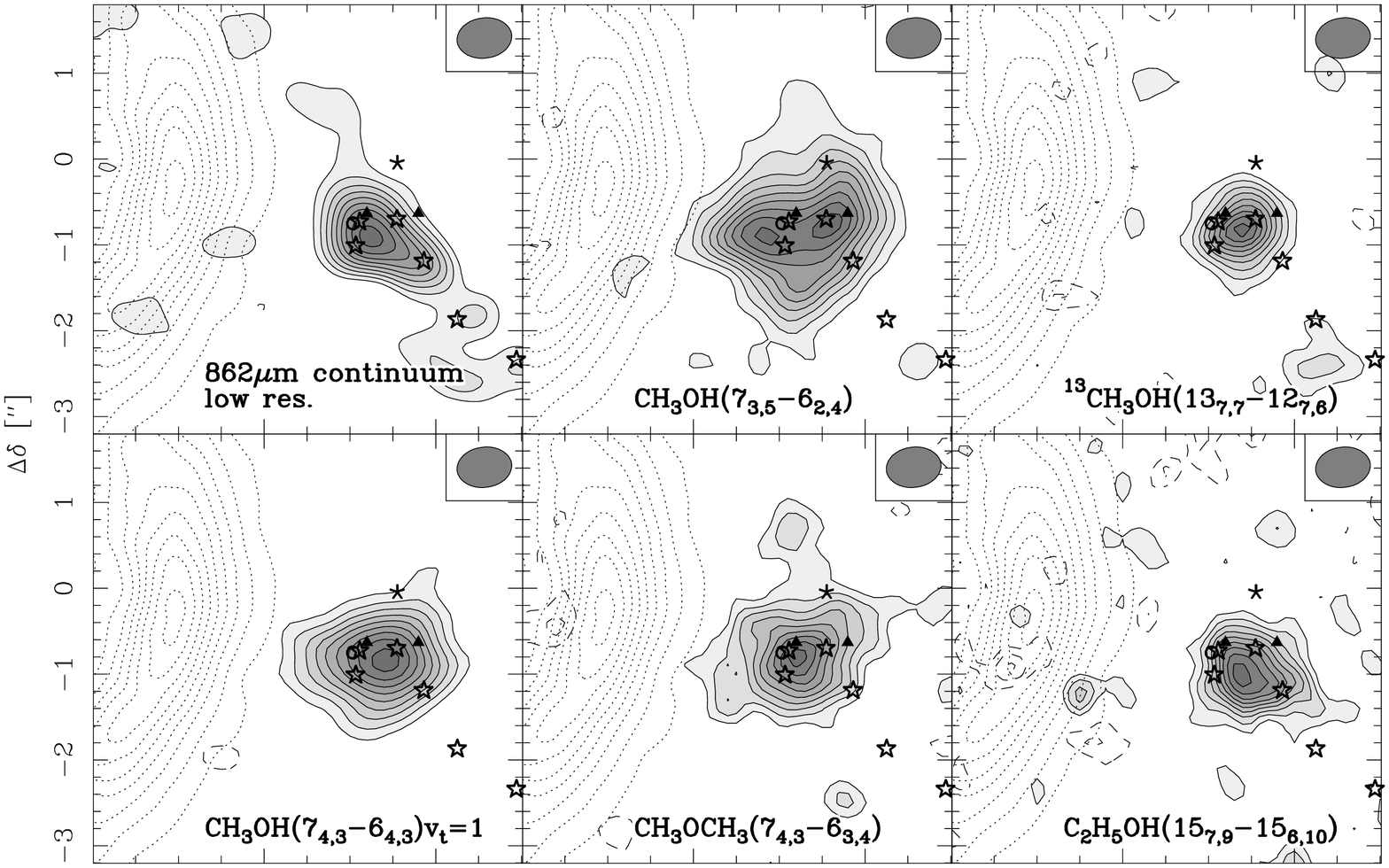}\\
\includegraphics[angle=-90,width=18cm]{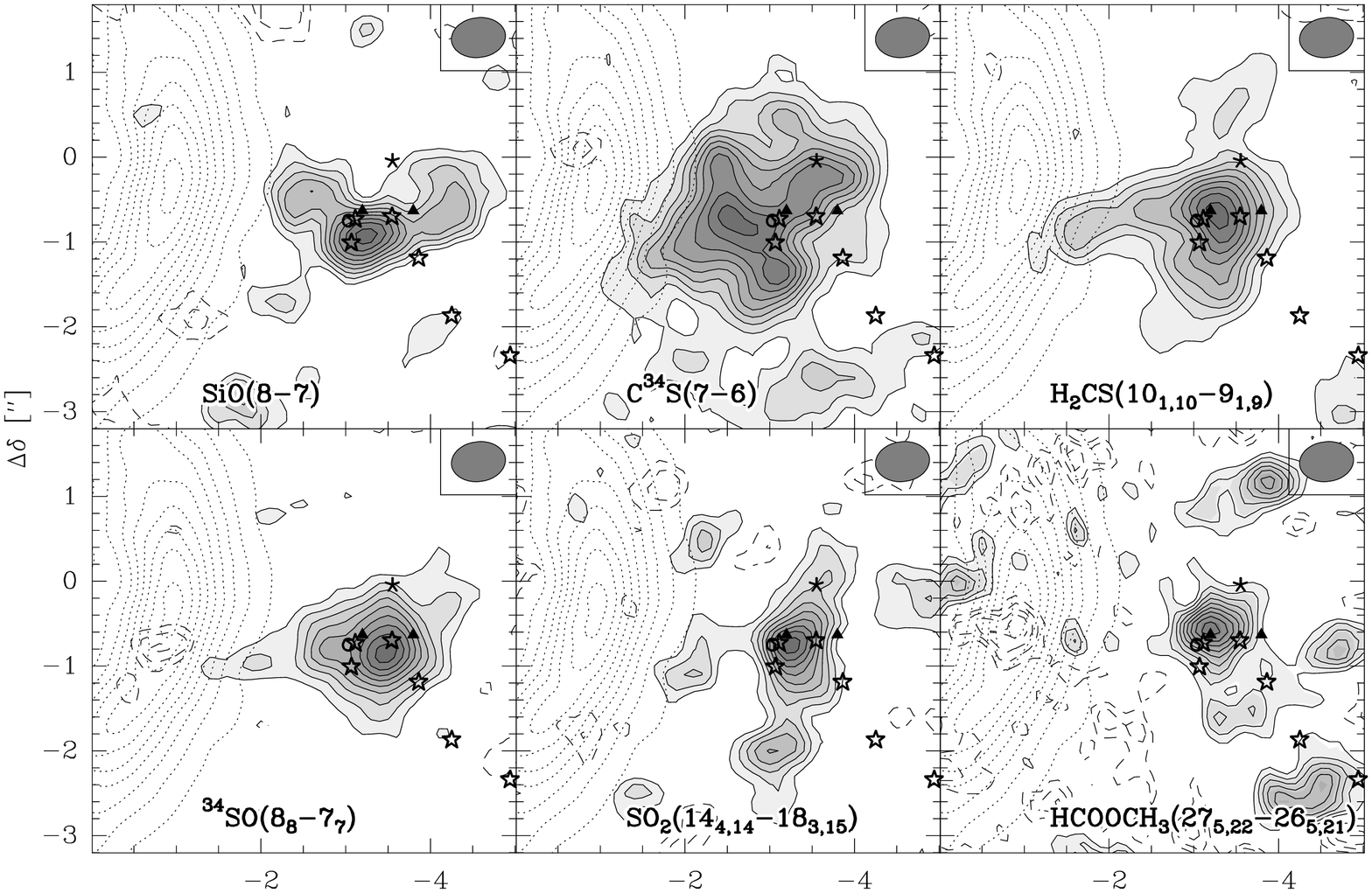}\\
\end{center}
\caption{Compilation of integrated line images (and submm continuum at
  the same spatial resolution) always shown in grey-scale with
  contours and labeled at the bottom of each panel. The dashed
  contours show negative features due to missing short spacings. The
  contouring is done from $\pm 15$ to $\pm$95\% (step $\pm$10\%) of
  the peak emission of each image, respectively. Peak fluxes
  $S_{\rm{peak}}$, rms and integrated velocity ranges for each image
  are given in Table \ref{rms}. The dotted contours again show the
  UCH{\sc ii} region and the stars mark the submm continuum peaks from
  Figure \ref{continuum}. The offsets on the axes are relative
    to the phase center.}
\label{lineimages}
\end{figure*}

\begin{figure*}[htb]
\begin{center}
\includegraphics[angle=-90,width=18cm]{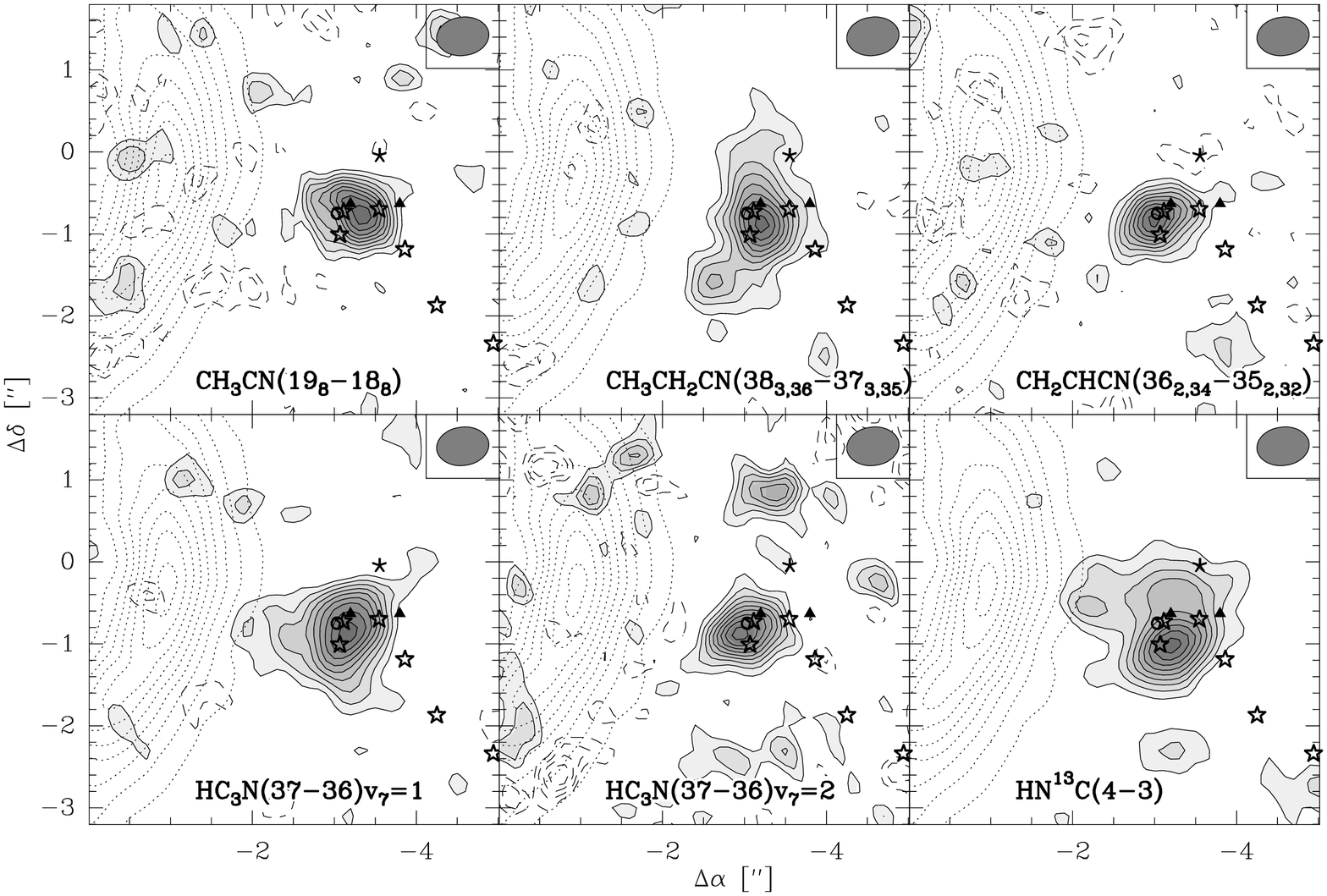}
\end{center}
\caption{Continued Figure \ref{lineimages}.}
\label{lineimages2}
\end{figure*}

Figures \ref{lineimages} and \ref{lineimages2} now present integrated
images of the various detected species, isotopologues and
vibrationally excited lines. For comparison, Figure \ref{lineimages}
also shows the submm continuum emission reduced with the same degraded
spatial resolution as the line images.  All images show emission in
the vicinity of the hot molecular core and no emission toward the
associated UCH{\sc ii} region. However, the morphology varies
significantly between many of the observed molecular line maps. The
molecular emission is largely confined to the central region of the
main four submm continuum peaks, and we do not detect appreciable
molecular emission toward the continuum peaks 5 and 6. Reducing the
submm continuum data with the same spatial resolution as the line
images, the four submm peaks are smoothed to a single elongated
structure peaking close to the submm peak submm1 (Fig.~\ref{lineimages},
top-left panel).  The ground state CH$_3$OH emission is relatively
broadly distributed with two peaks in east-west direction, and one may
associate one with the submm peaks 1 and 2 and the other with the
submm peak submm3, but most other maps show on average one spectral line
peak somewhere in the middle of the 4 main submm continuum peaks,
similar to the lower-resolution submm continuum map.

However, there are also a few species which significantly deviate from
this picture and show a different spatial morphology. For example SiO
is more extended in north-east south-west direction likely due to a
molecular outflow (\S \ref{outflow}). Also interesting is the emission
from C$^{34}$S which lacks emission around the central four submm
peaks but is stronger in the interface region between the hot
molecular core and the UCH{\sc ii} region (\S \ref{c34s}).
Furthermore, there are a few spectral line maps -- mainly those from
likely optically thin lines (HCOOCH$_3$, HN$^{13}$C), highly excited
lines (CH$_3$CHCN) and vibrationally excited lines (CH$_3$OH
$v_t=1,2$, HC$_3$N $v_7=1,2$) -- which show their emission peaks
concentrated toward the main submm peak submm1 (\S \ref{disk}).

Previous lower-resolution ($\sim 10''$) molecular line observations
revealed strong CH$_3$OH emission toward the H$_2$O maser feature
approximately $4''$ south-west of the hot core peak
(Fig.~\ref{continuum}, \citealt{hofner1996,pratap1999}). A little bit
surprising, we do not detect any CH$_3$OH emission (nor any other
species) toward that south-western position, even when imaged at low
angular resolution using only the compact configuration data
(therefore, we do not cover that position in Figures \ref{lineimages}
and \ref{lineimages2}). \citet{pratap1999} discuss mainly two
possibilities to explain this discrepancy: Either their observed
specific CH$_3$OH$(8_0-7_1)$ line is a weak maser and we do not cover
any comparable CH$_3$OH line, or the emission covered by the
lower-resolution data is relatively extended and filtered out by our
observations. As discussed in the previous section, the shortest
baseline of our observations was $\sim$16\,m, implying that we are not
sensitive to any scales $> 12''$. Since the CH$_3$OH emission in
\citet{pratap1999} is slightly resolved by their synthesized beam of
$12.6''\times 9.8''$, it is unlikely that we would have filtered out
all emission. However, among the many observed CH$_3$OH lines (Table
\ref{linelist}), some have similar excitation temperatures of the
order 80\,K as the line observed by \citet{pratap1999}, and we would
expect to detect thermal emission from these lines as well. Therefore,
our non-detection of CH$_3$OH emission toward the south-western H$_2$O
maser position supports rather their suggested scenario of weak
CH$_3$OH maser emission in the previously reported observations
\citep{pratap1999}.

\begin{figure}[htb]
\includegraphics[angle=-90,width=8.7cm]{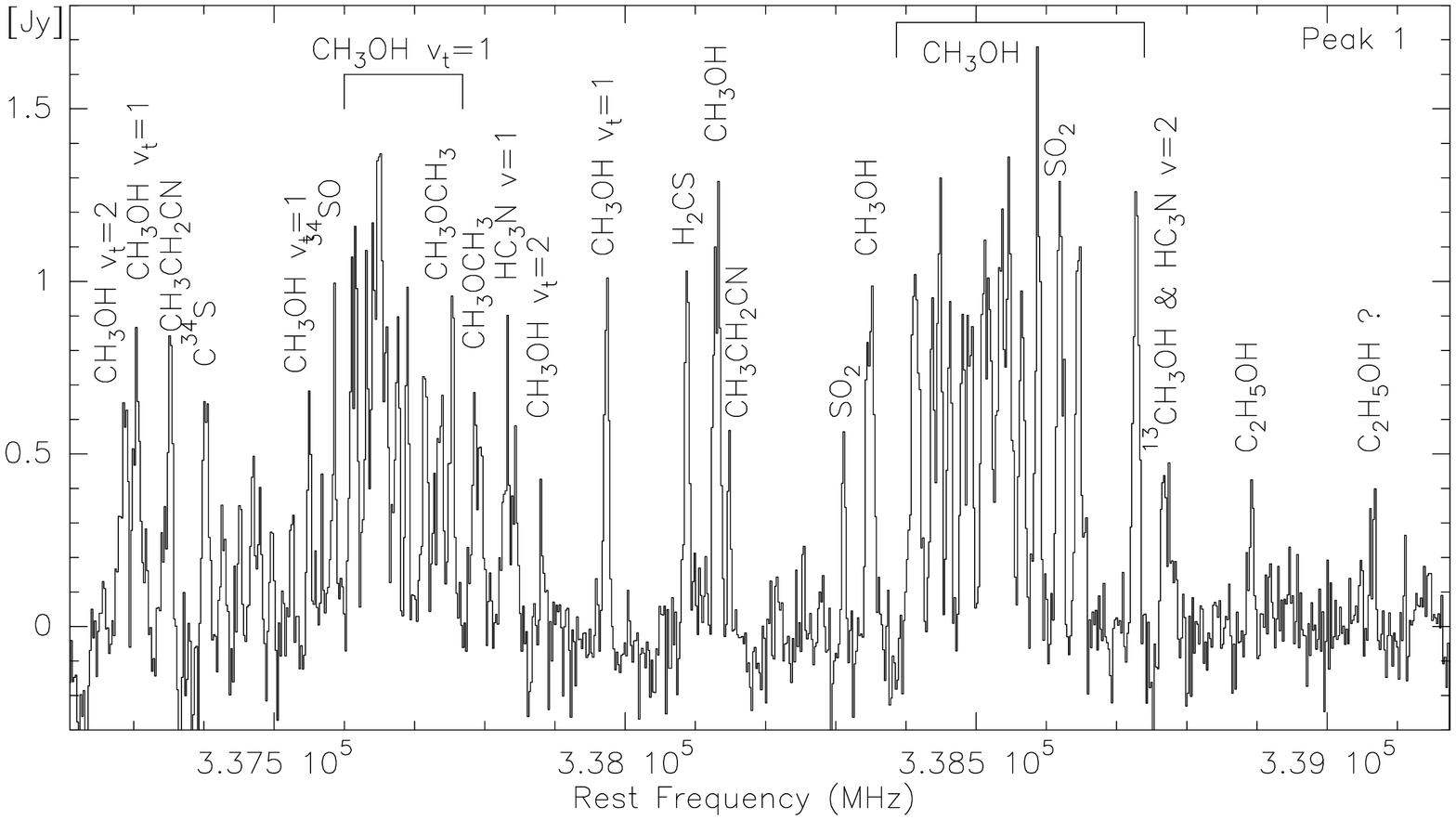}\\
\includegraphics[angle=-90,width=8.7cm]{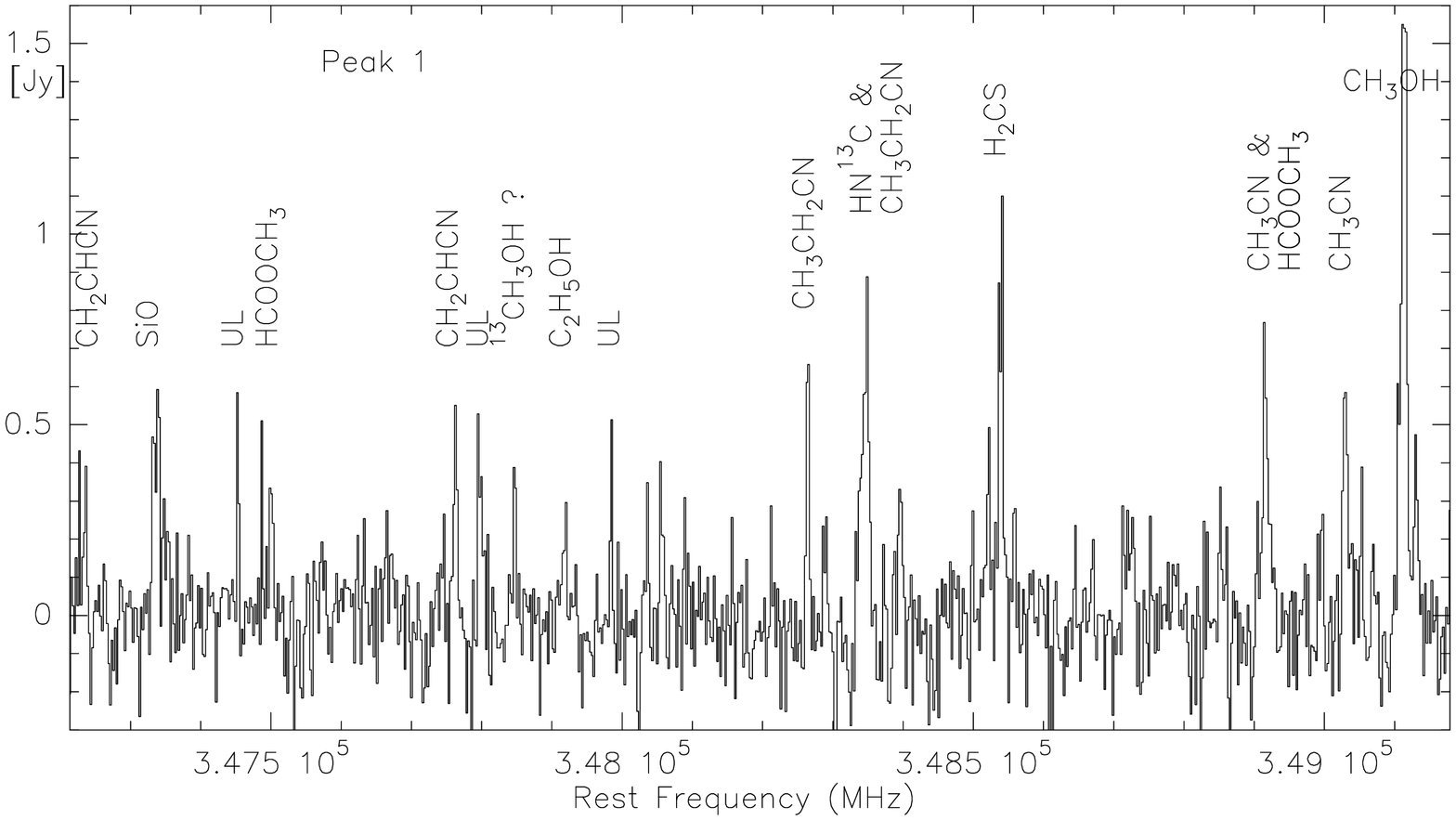}
\caption{Lower and upper sideband spectra extracted toward the submm1.
  The spatial resolution of these data is $0.64''\times 0.47''$. The
  main line identifications are shown in both panels.}
\label{spectra}
\end{figure}

\begin{table}[htb]
\caption{Detected molecular species}
\begin{tabular}{lrr}
\hline \hline
Species & Isotopologues & Vibrational states \\
\hline
CH$_3$OH & $^{13}$CH$_3$OH & CH$_3$OH, $v_t=1,2^a$ \\
CH$_3$OCH$_3$ \\
C$_2$H$_5$OH\\
SiO\\
   & C$^{34}$S \\
H$_2$CS\\
   & $^{34}$SO \\
SO$_2$\\
HCOOCH$_3$\\
CH$_3$CN\\
CH$_3$CH$_2$CN\\
CH$_3$CHCN \\
   & & HC$_3$N, $v_7=1,2$\\
   & HN$^{13}$C\\
\hline \hline
\multicolumn{3}{l}{\footnotesize $^a$ The detection of this CH$_3$OH $v_t=2$ line is}\\
\multicolumn{3}{l}{\footnotesize doubtful since other close $v_t=2$ lines with}\\
\multicolumn{3}{l}{\footnotesize similar excitation temperatures were not detected.}\\
\end{tabular}
\label{species}
\end{table}

\subsection{Molecular outflow emission}
\label{outflow}

\begin{figure}[htb]
\includegraphics[angle=-90,width=8.7cm]{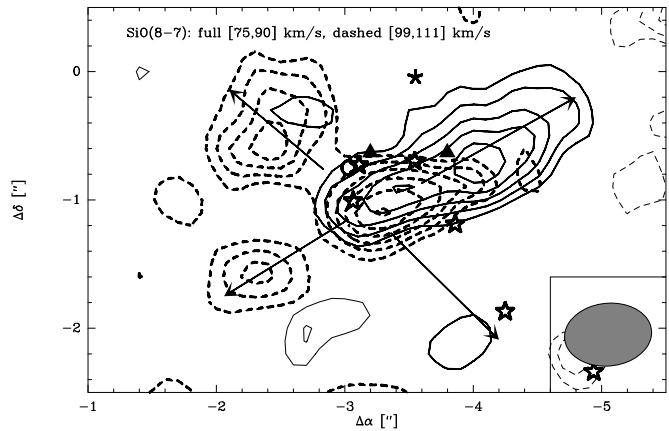}
\caption{SiO(8-7) outflow map. The full and dashed contours are
  integrated over the blue- and redshifted SiO emission as shown in
  the figure. The contouring starts at $\pm 2\sigma$ and continues in
  $\pm 1\sigma$ steps (thick contours positive, thin contours
  negative). The $1\sigma$ values for the blue- and red-shifted images
  are 48 and 46\,mJy\,beam$^{-1}$, respectively. The markers are the
  same as in the previous images, the synthesized beam of $0.68''
  \times 0.49''$ is shown at the bottom right, and the arrows guide
  the eye for the potential directions of the two discussed outflows.
  The offsets on the axes are relative to the phase center.}
\label{sio}
\end{figure}

The SiO(8-7) spectrum spans a large range of velocities from $\sim$75
to $\sim$111\,km\,s$^{-1}$. Integrating the blue- and red-shifted
emission, one gets the outflow map presented in Fig.~\ref{sio}. The
elongated north-west south-east structure is consistent with the
previously proposed outflow by \citet{gibb2004}. The additional red
feature north-east of the central hot core region makes the
interpretation ambiguous: If the north-west south-east outflow is a
relatively highly collimated jet, then the north-eastern red feature
could be attributed to an additional outflow leaving the core in
north-east south-west direction. The blue wing of that potential
second outflow would not detected in our data. However, since we are
filtering out any larger-scale emission, it is also possible that the
red SiO features south-east and north-east of the main core are part
of the same wide-angle outflow tracing potentially the limb-brightened
cavity walls. In this scenario, our observations would miss part of
the blue-shifted wide-angle outflow lobe.  With the current data, it
is difficult to clearly distinguish between the two scenarios.
However, comparing the elongated blue-shifted SiO(8--7) data with the
previous north-west south-eastern outflow observed in H$_2$S by
\citet{gibb2004}, it appears that this is the most likely direction of
the main outflow of the region. Therefore, the multiple outflow
scenario appears more likely for the hot core in G29.96$-$0.02.  The
lower resolution SiO(2--1) observation by \citet{maxia2001} are also
consistent with this scenario. Based on these data, we cannot
conclusively say which of the submm continuum sources submm1 to submm4
contribute to driving the outflows.

\section{Discussion}

\subsection{The formation of a proto-Trapezium system?}

The four main submm continuum peaks are located within a projected
area of $7800\times 7800$\,(AU)$^2$ on the sky. The projected
separation $\Delta \theta$ between individual sub-sources varies
between 1800\,AU (peaks 1 and 2) and 5400\,AU (peaks 1 and 4, see
Table \ref{separation}). Could the four central submm peaks be the
predecessors of a future Trapezium system? Trapezia are defined as
non-hierarchical multiple systems of three or more stars where the
largest projected separation between Trapezia members should not
exceed the smallest projected separation by a factor of 3
\citep{sharpless1954,ambartsumian1955,abt2000}. This criterion is
satisfied by the four submm peaks at the center of the G29.96$-$0.02
hot core.  The 14 optically identified Trapezia discussed by
\citet{abt2000} have mean radii to the furthest outlying member of
$\sim 4\times 10^4$\,AU, with the largest radius of $\sim
5.4\times10^5$\,AU ($\sim$2.6\,pc), the approximate dimension of an
open cluster. Therefore, the protostellar projected separations of the
tentative proto-Trapezium candidate in G29.96$-$0.02 are significantly
smaller than in typical optically visible Trapezia systems.  A similar
small size for a candidate Trapezium system has recently been reported
for the multiple system in W3IRS5 \citep{megeath2005}.

\begin{table}[htb]
\caption{Spatial separation}
\begin{tabular}{lrr}
\hline \hline
Pair& $\Delta \theta$ & $\Delta x$ \\
    & [$''$]     & [AU] \\
\hline 
1-2 & 0.3 & 1800 \\
1-3 & 0.5 & 3000 \\
1-4 & 0.9 & 5400 \\
2-3 & 0.6 & 3600 \\
2-4 & 0.8 & 4800 \\
3-4 & 0.6 & 3600 \\
\hline \hline
\end{tabular}
~\\
\footnotesize{The numbers in column 1 correspond to the numbers of the submm peaks.}
\label{separation}
\end{table}

The small sizes of the proto-Trapezia in G29.96$-$0.02 and W3IRS5 may
be attributed to their youth. During their upcoming evolution, these
young system will expel most of the surrounding gas and dust envelope
via the protostellar outflows and strong uv-radiation. Therefore, the
whole gravitational potential of the system will decrease and the
kinetic energy may dominate. Systems with positive total energy will
globally expand and will eventually be observable as a larger-scale
optical Trapezia systems \citep{ambartsumian1955}.

With the given data it is hard to estimate how massive the expected
Trapezia stars are and will finally be at the end of their formation
processes. The integrated hot core luminosity is estimated to be $\sim
10^5$\,L$_{\odot}$ \citep{cesaroni1994,olmi2003}, in contrast to the
integrated luminosity of the whole region measured by the large IRAS
beam of $\sim 10^6$\,L$_{\odot}$. Producing $10^5$\,L$_{\odot}$
requires either an O7 star or a few stars of comparable but lower
masses. Nevertheless, the numbers imply that this Trapezium system
should form at least one or more massive stars.  Although the gas
masses we derived from our dust continuum data (Table \ref{submmcont})
are relatively low, that does not necessarily imply that their mass
reservoir is restricted to these gas masses because it is possible
that they may accrete additional gas from the larger-scale envelope
that is filtered out by our observations. This scenario is predicted
by the competitive accretion model for massive star formation (e.g.,
\citealt{bonnell2004}). The fact that the gas masses we find for the
four strongest submm sources are all similar allows to speculate that
they may form about similar mass stars in the end, however, this
cannot be proven by these data in more detail.

Assuming that the projected size of the potential proto-Trapezium
system in G29.96$-$0.02 of approximately 7800\,(AU)$^2$ resembles a
3-dimensional sphere of radius $\sim$3900\,AU, we can estimate the
current protostellar volume density of the region to approximately
$1.4\times 10^5$ protostars per cubic pc. This number is larger than
typical stellar densities in young clusters of the order $10^4$ stars
per cubic pc \citep{lada2003}, but it is still below the extremely
high (proto)stellar densities required for protostellar merger models
of the order $10^6$ to $10^8$ stars per cubic pc
\citep{bonnell1998,bonnell2004,stahler2000,bally2005}.

Although we have not yet observed the extremely high (proto)stellar
densities predicted by the coalescence scenario, as soon as we observe
massive star-forming regions with a spatial resolution $\leq
4000$\,AU, we begin to resolve multiplicity and potential
proto-Trapezia (see also the recent observations of NGC6334I and I(N)
by \citealt{hunter2006}). Furthermore, this (proto)stellar density may
even be a lower limit, since we observe only a two-dimensional
projection and are additionally sensitivity limited to masses $\geq
2.1$\,M$_{\odot}$ (corresponding to the 3$\sigma$ flux limit of
63\,mJy\,beam$^{-1}$ at the assumed temperature of 100\,K).  Higher
spatial resolution has so far always increased the observed
(proto)stellar densities, and it is possible that in the future we may
reach the $10^6$ requirement for merging to play a role. However, it
is also important to get better theoretical predictions of potential
merger signatures that observers could look for.

\subsection{Various episodes of massive star formation?}

It is interesting to note that the previously identified mid-infrared
source \citep{debuizer2002} is offset from the submm continuum peaks.
Although the mid-infrared astrometry is usually relatively uncertain,
the association of the mid-infrared peak with class {\sc ii} CH$_3$OH
maser emission with an absolute positional uncertainty of only 30\,mas
\citep{minier2001} is indicative that the offset may be real.
Combining the facts that we find within a small region of only
$\sim$20000\,AU ($\sim$0.1\,pc) at least three different regions of
massive star formation -- the UCH{\sc ii} region, the mid-infrared
source, and the submm continuum sources -- indicates that not all
massive stars within the same evolving cluster are coeval but that
sequences of massive star formation may take place even on such small
spatial scales.

\subsection{Carbon mono-sulfide C$^{34}$S}
\label{c34s}

One of the most striking spectral line maps is from the rare carbon
mono-sulfide isotopologue C$^{34}$S(7--6). Its emission peak is not
toward the hot core nor any of the submm continuum peaks, but largely
east of it in the interface region between the submm continuum peaks
and the UCH{\sc ii} region. Hence, one likes to understand why the
C$^{34}$S emission is that weak toward the hot core region and that
strong at the hot core/UCH{\sc ii} region interface.

CS usually desorbs from dust grains at moderate temperatures of a few
10\,K, hence it should be observable relatively early in the evolution
of a growing hot molecular core (e.g., \citealt{viti2004}). From
100\,K upwards H$_2$O is released from grains, then it forms OH
molecules, and the OH can react with S to SO and SO$_2$ (e.g.,
\citealt{charnley1997}). Therefore, the initial high CS abundances
should decrease with time while the SO and SO$_2$ are expected to
increase with time (e.g., \citet{wakelam2005b}). As shown in Figure
\ref{lineimages}, $^{34}$SO peaks toward the hot core where the
derived CH$_3$OH temperatures exceed the H$_2$O evaporation
temperature (see \S \ref{temperature} and Fig.~7b, potentially
validating this theoretical prediction. According to such chemical
models, the hot core G29.96$-$0.02 should have a chemical age of at
least a few times 10$^4$ years.

The strong C$^{34}$S emission in the hot core/UCH{\sc ii} interface
region may be explained in the same framework. In the molecular
evolution scheme outlined above, one would expect low C$^{34}$S
emission toward the hot core with a maybe symmetrical increase
further-out. In the case of the G29.96$-$0.02 hot core, we have the
decrease toward the center, but the emission rises only toward the
east, north and west with the strongest increase in the eastern hot
core/UCH{\sc ii} region interface. If one compares the C$^{34}$S
morphology in Figure \ref{lineimages} with the temperature
distribution in Figure 7b, one finds the lowest CH$_3$OH
temperatures right in the vicinity of the C$^{34}$S emission peaks,
adding further support to the proposed chemical picture.

Extrapolating this scenario to other molecules, it indicates that
species which are destroyed by H$_2$O, e.g., molecular ions such as
HCO$^+$ or N$_2$H$^+$ (e.g., \citealt{bergin1998}), are no good probes
of the inner regions of hot molecular cores.

\subsection{Temperature structure}
\label{temperature}

\citet{leurini2004,leurini2007} investigated the diagnostic properties
of methanol over a range of physical parameters typical of high-mass
star-forming regions. They found that the ground state lines of
CH$_3$OH are mainly tracers of the spatial density of the gas,
although at submillimeter wavelengths high $k$ transitions are also
sensitive to the kinetic temperature. However, in hot, dense regions
such as hot cores, the effects of infrared pumping on the level
populations due to the thermal heating of the dust is not negligible,
but mimic the effect of collisional excitation.  For the ground state
line, \citet{leurini2007} found that it is virtually impossible to
distinguish between IR pumping and pumping by collisions, as both
mechanisms equally populate the $v_t=0$ levels. On the other hand, the
vibrationally or torsionally excited lines have very high critical
densities ($ 10^{10}$--10$^{11}$ cm$^{-3}$) and high level energies
($T\ge 200$~K). They are difficult to be populated by collisions and
trace the IR field instead.

To study the physical conditions of the gas around the main continuum
peaks in G29.96--0.02, we analyzed only the emission coming from the
$v_t=1$ lines, as their optical depth is lower than for the ground
state, and their emission is confined to the gas around the dust
condensations, while the $v_t=0$ transitions are more extended and can
be affected by problems of missing flux.  We first fitted the methanol
emission of the $v_t=1$ lines (see Fig.~\ref{ch3ohfit}) towards the
peak position, using the method described by
\citet{leurini2004,leurini2007} that is based on an LVG analysis and
includes radiative pumping \citep{leurini2007}. The continuum emission
derived in $\S$3.1 was used in the calculations to solve the equations
for the level populations.  The two main dust condensations submm1 and
submm2 fall in the beam of the line data; however, we assumed that the
emission is coming from only one component, which is more extended
than our beam, and derived a CH$_3$OH column density averaged over the
beam of $4\times 10^{17}$~cm$^{-2}$. The corresponding methanol
abundance, relative to H$_2$ is of the order of $10^{-7}$, typical of
hot core sources.  Since the emission from the $v_t=1$ lines is
optically thin for this column density, and also at higher values, we
consider this approach valid. The temperature derived toward the line
peak is 340~K. This corresponds to our best fit model, but from a
$\chi^2$ analysis we can only infer a low limit of $\sim$220~K for the
temperature of the gas.  Since lines are optically thin, the
degeneracy between kinetic temperature and column density is not
solved, and the model delivers good fit to the $v_t=1$ lines for lower
or higher temperatures by adjusting the methanol column density.
However, the low temperature solutions ($T_{kin}$=100--200~K) need
high methanol abundances relative to H$_2$($\sim 10^{-6}$), which can
be hardly found at these temperatures. Moreover, lines are optically
thick for these column densities, and the assumption of our analysis
is not valid anymore.

We also investigated the line ratio between several $v_t=1$ lines at
the column density derived for the main position, to find the best
temperature diagnostic tool among the methanol lines and derive a
temperature map of the region. We found that the line ratios with the
blend of lines at $\sim 337.64$~GHz increase with the temperature of
the gas (Fig.~7a). However, the blending of several
transitions together complicates the use of such diagnostic. In
Fig.~7b, we show the map of the line ratio between the
$7_{1,6}\to 6_{1,5}$-$E$~$v_t=1$ at 337.708~GHz and the blend between
the $7_{1,7}\to 6_{1,6}$-$E$~$v_t=1$ at 337.642~GHz and $7_{0,7}\to
6_{0,6}$-$E$~$v_t=1$ at 337.644~GHz. Since line intensities do not
simply add up, we did not correct for the overlapping between the two
transitions. Two other lines, the $7_{4,3}\to 6_{4,2}$-$E$~$v_t=1$ and
the $7_{5,3}\to 6_{4,2}$-$E$~$v_t=1$, are also very close in
frequency. This is seen in the linewidth of the blending, which is
wider than for the other lines. Therefore, we considered only half of
the channels of the blending at 337.64~GHz in our line ratio analysis.
From the ratio-map in Fig.~7b, submm1, submm2 and submm3 of Table~1 show
high temperatures (T$\ge 300$~K), while relatively low temperature gas
(T$\sim 100$~K) is found at R.A. [J2000]=$18^h46^m03^s.818$ Dec.
[J2000]= $-02^\circ 39'22''.14$, close to a secondary peak of many
ground state lines of methanol (Fig.~\ref{lineimages}). The temperature
then decreases towards submm4.  The increase in the line ratio towards
the south-east and north is probably not true, but due to the poor
signal to noise ratio in these areas.  Changes in the column densities
along the area may affect our results.

\begin{figure}[htb]
\caption{\small Spectrum of the $7_{k_a,k_b} \to
  6_{k_a,k_b-1}$~$v_t=1$ methanol band toward the main dust
  condensation.  Overlaid in black is the synthetic spectrum resulting
  from the fit.}
\label{ch3ohfit}
\end{figure}

\begin{figure}[h]
\centering
\caption{\small {\bf a:} Modeled line ratio between the $7_{1,6}\to
  6_{1,5}$-$E$~$v_t=1$ line and the $7_{1,7}\to 6_{1,6}$-$E$~$v_t=1$
  transitions, as function of the temperature. {\bf b:} Map of the
  line ratio between the same transitions in the inner region around
  the peaks.  The white stars mark the positions of the dust peaks;
  the white dashed contours show the values of the line ratio from
  $\sim150$ to $\sim350$~K, which correspond to levels from 0.3 to 0.7
  in step of 0.1 in the map. The solid black contours show the
  continuum emission smoothed to the resolution of the line data (from
  0.2 to 0.4 Jy/beam in step of 0.05). The offsets on the axes are
  relative to the phase center.}
\end{figure}

\subsection{Tracing rotation toward the massive cores}
\label{disk}

At the given lower spatial resolution of the spectral line data
compared to the submm continuum, we cannot resolve the four submm
peaks well. However, one of the aims of such multi-line studies is to
identify spectral lines that trace the massive protostars and that are
potentially associated with massive disk-like structures. Such lines
may then be used for kinematic gas studies of rotating gas envelopes,
tori or accretion disks. Therefore, we analyzed the data-cubes
searching for velocity structures indicative of any kind of rotation.
In the large majority of spectral lines, this was not successful and
we could mostly not identify coherent velocity structure.  While
chemical and temperature effects (\S \ref{c34s} \& \ref{temperature})
may be responsible for parts of that, the large column densities
derived in \S \ref{continuumtext} imply also large molecular line
column densities and hence large optical depths.  Therefore, many of
the observed lines are likely optically thick tracing only outer gas
layers of the hot molecular core not penetrating down to the deeply
embedded protostars. Furthermore, many molecules would not only be
excited in the central rotating disk-like structures but also in the
surrounding envelope and maybe the outflow.  Hence, disentangling the
different components observationally remains a challenging task.

\begin{figure}[htb]
\caption{Moment 1 maps of HN$^{13}$C(4--3) (top) and
  HC$_3$N(37--36)$v_7=1$ (bottom). The markers are the same as in the
  previous images, and the synthesized beam of $0.68'' \times 0.49''$
  is shown at the bottom left. The offsets on the axes are relative to
  the phase center.}
\label{mom1}
\end{figure}

The major exceptions are the molecular lines of the rare
isotopologue of hydrogen isocyanide HN$^{13}$C(4--3) with a low
excitation temperature of only 42\,K, and the vibrationally excited
line of cyanoacetylene HC$_3$N(37--36)$v_7=1$ with a higher excitation
temperature of 629\,K (Fig.~\ref{mom1}). In both cases we find a
velocity gradient across the main submm peak submm1 with a position angle of
$\sim 45^{\circ}$ from north. This is approximately perpendicular to
the molecular outflow discussed in \S\ref{outflow} and by
\citet{gibb2004}. Interestingly, \citet{gibb2004} also find a similar
velocity gradient in their central velocity channels of H$_2$S. The
previously reported NH$_3$ and CH$_3$CN velocity gradients in
approximately east-west direction \citep{cesaroni1998,olmi2003} have
been observed with slightly lower spatial resolution and are
consistent with our data as well.

Our observations as well as previous work in the literature suggest
that the G29.96$-$0.02 hot core exhibits a velocity gradient in the
dense gas in approximately north-east south-west direction
perpendicular to the molecular outflow observed at larger scales.
Based on the HN$^{13}$C(4--3) map, the diameter of this structure is
$\sim 1.6''$ corresponding to radius of $\sim$4800\,AU.  Since this
emission encompasses not only the submm peak submm1 but also the
submm2 and submm3, it is not genuine protostellar disk as often
observed in low-mass star-forming regions.  The velocity structure
does not resemble Keplerian rotation and may hence be due to some
larger-scale rotating envelope or torus that could transform into a
genuine accretion disks at smaller still unresolved spatial scales
\citep{cesaroni2006}. Additional options to explain such a velocity
gradient may be (a) interaction with the 2nd outflow in
north-east--south-western direction, (b) interaction with the
expanding UCH{\sc ii} region, and (c) global collapse like recently
proposed for NGC2264 \citep{peretto2007}. While we cannot exclude (a)
and (b), option (c) of a globally collapsing core appears particularly
interesting because combining rotation and collapse would result in an
inward spiraling kinematic structure, potentially similar to the
models originally proposed for rotating low-mass cores (e.g.,
\citealt{ulrich1976,terebey1984}). Recent hydrodynamic simulations by
\citet{dobbs2005} and \citet{krumholz2006b} as well as analytic
studies by \citet{kratter2006} find fragmentation and star formation
within the massive disks forming early in the collapse process of
high-mass cores. This would be consistent with the found three
sub-sources (submm1 to submm3) within the HN$^{13}$C/HC$_3$N
structure. However, on a cautionary note it needs to be stressed that
the collapse/rotation scenario is far from conclusive, and that the
outflow and/or UCH{\sc ii} region can potentially influence the
observed velocity pattern as well. It remains puzzling that only these
two lines exhibit the discussed signatures whereas all the other
spectral lines in our setup do not.

\section{Conclusions and Summary}

The new 862\,$\mu$m submm continuum and spectral line data obtained
with the SMA toward G29.96$-$0.02 at sub-arcsecond spatial resolution
resolve the hot molecular core into several sub-sources. At an angular
resolution of $0.36''\times 0.25''$, corresponding to linear scales of
$\sim$1800\,AU, the central core contains four submm continuum peaks
which resemble a Trapezium-like multiple system at a very early
evolutionary stage.  Assuming spherical symmetry for the hot core
region, the protostellar densities are high of the order $1.4\times
10^5$ protostars per pc$^3$.  However, these protostellar densities
are still below the required values between $10^6$ to $10^8$
protostars/pc$^3$ to make coalescence of protostars a feasible
process. Derived H$_2$ column densities of the order a few
$10^{24}$\,cm$^{-2}$ imply visual extinctions of a few 1000.  The
existence of three sites of massive star formation in different
evolutionary stages within a small region (the UCH{\sc ii} region, the
mid-infrared source, and the submm continuum sources) indicates that
sequences of massive star formation may take place within the same
evolving massive protocluster.

The 4\,GHz of observed bandpass reveal a plethora of approximately 80
spectral lines from 18 molecular species, isotopologues or
vibrationally excited lines. Only about 5\% of the spectral lines
remain unidentified. Most spectral lines peak toward the hot molecular
core, while a few species also show more extended emission, likely due
to molecular outflows and chemical differentiation. The CH$_3$OH line
forest allows us to investigate the temperature structure in more
detail. We find hot core temperatures $\geq 300$\,K and decreasing
temperature gradients to the core edges. The SiO(8-7) observations
confirm a previously reported outflow \citet{gibb2004} in north-west
south-east direction with a potential identification of a second
outflow emanating approximately in perpendicular direction.
Furthermore, C$^{34}$S exhibits a peculiar morphology being weak
toward the hot molecular core and strong in its surroundings,
particular in the UCH{\sc ii}/hot core interface region.  The
C$^{34}$S deficiency toward the hot molecular core may be explained by
time-dependent chemical desorption from grains, where the C$^{34}$S
desorbs early, and later-on after H$_2$O desorbs from grains forming
OH, the sulphur reacts with the OH to form SO and SO$_2$.

Furthermore, we were interested in identifying the best molecular line
tracers to investigate the kinematics and potential disk-like
structures in such dense and young massive star-forming regions.  Most
spectral lines do not exhibit any coherent velocity structure.  A
likely explanation for this uncorrelation between molecular line peaks
and submm continuum peaks is that many spectral lines may be optically
thick in such high-column-density regions, and that additional
chemical evolution and temperature effects complicate the picture.
Furthermore, many molecules are excited in various gas components
(e.g., disk, envelope, outflow), and it is often observationally
difficult to disentangle the different contributions properly. There
are a few exceptions of optically thin and vibrationally excited lines
that apparently probe deeper into the core tracing submm1 better than
other transitions.  Investigating the velocity pattern of these
spectral lines, we find for some of them a velocity gradient in the
north-east south-west direction perpendicular to the molecular
outflow. Since the spatial scale of this structure is relatively large
($\sim$4800\,AU) comprising three of the central protostellar sources,
and since the velocity structure is not Keplerian, this is not a
genuine Keplerian accretion disk. While these data are consistent with
a larger-scale toroid or envelope that may rotate and/or globally
collapse, we cannot exclude other explanations, such as that the
influence of the outflow(s) and/or expanding UCHII region produces the
observed velocity pattern.  In addition to this, these data confirm
previous findings that the high column densities, the large optical
depths of the spectral lines, the chemical evolution, and the
different spectral line contributions from various gas components make
it very difficult to identify suitable massive accretion disk tracers,
and hence to study this phenomenon in a more statistical fashion.
(e.g., \citealt{beuther2006c})

\begin{acknowledgements} 
  We like to thank Peter Schilke and Sebastian Wolf for many
  interesting discussions about related subjects. We also thank the
  anonymous referee whose comments helped improving the paper.
  H.B.~acknowledges financial support by the Emmy-Noether-Program of
  the Deutsche Forschungsgemeinschaft (DFG, grant BE2578).
\end{acknowledgements}


\begin{longtable}[htb]{lrr|rrr}
\caption{Line parameters}\\
\hline
\hline
Freq. & Line & $E_u$ & Freq. & Line & $E_u$\\
GHz &      & K & GHz &      & K\\
\hline
337.279 & CH$_3$OH$(7_{2,5}-6_{2,4})$E($v_t$=2)$^a$    & 727 & 338.409 & CH$_3$OH$(7_{0,7}-6_{0,6})$A                 & 65  \\
337.297 & CH$_3$OH$(7_{1,7}-6_{1,6})$A($v_t$=1)        & 390 & 338.431 & CH$_3$OH$(7_{6,1}-6_{6,0})$E                 & 254 \\
337.348 & CH$_3$CH$_2$CN$(38_{3,36}-37_{3,35})$        & 328 & 338.442 & CH$_3$OH$(7_{6,1}-6_{6,0})$A                 & 259 \\
337.397 & C$^{34}$S(7--6)                              & 65  &         & CH$_3$OH$(7_{6,2}-6_{6,1})$A$^-$             & 259 \\
337.421 & CH$_3$OCH$_3(21_{2,19}-20_{3,18})$           & 220 & 338.457 & CH$_3$OH$(7_{5,2}-6_{5,1})$E                 & 189 \\
337.446 & CH$_3$CH$_2$CN$(37_{4,33}-36_{4,32})$        & 322 & 338.475 & CH$_3$OH$(7_{5,3}-6_{5,2})$E                 & 201 \\
337.464 & CH$_3$OH$(7_{6,1}-6_{0,0})$A($v_t$=1)        & 533 & 338.486 & CH$_3$OH$(7_{5,3}-6_{5,2})$A                 & 203 \\
337.474 & UL                                           &     &         & CH$_3$OH$(7_{5,2}-6_{5,1})$A$^-$             & 203 \\
337.490 & HCOOCH$_3(27_{8,20}-26_{8,19})$E             & 267 & 338.504 & CH$_3$OH$(7_{4,4}-6_{4,3})$E                 & 153 \\
337.519 & CH$_3$OH$(7_{5,2}-6_{5,2})$E($v_t$=1)        & 482 & 338.513 & CH$_3$OH$(7_{4,4}-6_{4,3})$A$^-$             & 145 \\
337.546 & CH$_3$OH$(7_{5,3}-6_{5,2})$A($v_t$=1)        & 485 &         & CH$_3$OH$(7_{4,3}-6_{4,2})$A                 & 145 \\
        & CH$_3$OH$(7_{5,2}-6_{5,1})$A$^{-}$($v_t$=1)  & 485 &         & CH$_3$OH$(7_{2,6}-6_{2,5})$A$^-$             & 103 \\
337.582 & $^{34}$SO$(8_8-7_7)$                         & 86  & 338.530 & CH$_3$OH$(7_{4,3}-6_{4,2})$E                 & 161 \\
337.605 & CH$_3$OH$(7_{2,5}-6_{2,4})$E($v_t$=1)        & 429 & 338.541 & CH$_3$OH$(7_{3,5}-6_{3,4})$A$^+$             & 115 \\
337.611 & CH$_3$OH$(7_{6,1}-6_{6,0})$E($v_t$=1)        & 657 & 338.543 & CH$_3$OH$(7_{3,4}-6_{3,3})$A$^-$             & 115 \\
        & CH$_3$OH$(7_{3,4}-6_{3,3})$E($v_t$=1)        & 388 & 338.560 & CH$_3$OH$(7_{3,5}-6_{3,4})$E                 & 128 \\
337.626 & CH$_3$OH$(7_{2,5}-6_{2,4})$A($v_t$=1)        & 364 & 338.583 & CH$_3$OH$(7_{3,4}-6_{3,3})$E                 & 113 \\
337.636 & CH$_3$OH$(7_{2,6}-6_{2,5})$A$^-$($v_t$=1)    & 364 & 338.612 & SO$_2(20_{1,19}-19_{2,18})$                  & 199 \\
337.642 & CH$_3$OH$(7_{1,7}-6_{1,6})$E($v_t$=1)        & 356 & 338.615 & CH$_3$OH$(7_{1,6}-6_{1,5})$E                 & 86  \\
337.644 & CH$_3$OH$(7_{0,7}-6_{0,6})$E($v_t$=1)        & 365 & 338.640 & CH$_3$OH$(7_{2,5}-6_{2,4})$A                 & 103 \\
337.646 & CH$_3$OH$(7_{4,3}-6_{4,2})$E($v_t$=1)        & 470 & 338.722 & CH$_3$OH$(7_{2,5}-6_{2,4})$E                 & 87  \\
337.648 & CH$_3$OH$(7_{5,3}-6_{5,2})$E($v_t$=1)        & 611 & 338.723 & CH$_3$OH$(7_{2,6}-6_{2,5})$E                 & 91  \\
337.655 & CH$_3$OH$(7_{3,5}-6_{3,4})$A($v_t$=1)        & 461 & 338.760 & $^{13}$CH$_3$OH$(13_{7,7}-12_{7,6})$A        & 206 \\
        & CH$_3$OH$(7_{3,4}-6_{3,3})$A$^-$($v_t$=1)    & 461 & 338.769 & HC$_3$N$(37-36)v_7=2$                        & 525 \\
337.671 & CH$_3$OH$(7_{2,6}-6_{2,5})$E($v_t$=1)        & 465 & 338.886 & C$_2$H$_5$OH$(15_{7,8}-15_{6,19})$           & 162 \\
337.686 & CH$_3$OH$(7_{4,3}-6_{4,2})$A($v_t$=1)        & 546 & 339.058 & C$_2$H$_5$OH$(14_{7,7}-14_{6,8})$            & 150 \\
        & CH$_3$OH$(7_{4,4}-6_{4,3})$A$^-$($v_t$=1)    & 546 & 347.232 & CH$_2$CHCN$(38_{1,38}-37_{1,37})$            & 329 \\
        & CH$_3$OH$(7_{5,2}-6_{5,1})$E($v_t$=1)        & 494 & 347.331 &  $^{28}$SiO(8--7)                            & 75  \\
337.708 & CH$_3$OH$(7_{1,6}-6_{1,5})$E($v_t$=1)        & 489 & 347.446 & UL                                                 \\
337.722 & CH$_3$OCH$_3(7_{4,4}-6_{3,3})$EE             & 48  & 347.494 &  HCOOCH$_3(27_{5,22}-26_{5,21})$A            & 247 \\
337.732 & CH$_3$OCH$_3(7_{4,3}-6_{3,3})$EE             & 48  & 347.759 &  CH$_2$CHCN$(36_{2,34}-35_{2,32})$           & 317 \\
337.749 & CH$_3$OH$(7_{0,7}-6_{0,6})$A($v_t$=1)        & 489 & 347.792 & UL                                                 \\
337.778 & CH$_3$OCH$_3(7_{4,4}-6_{3,4})$EE             & 48  & 347.842 & UL                                                 \\
337.787 & CH$_3$OCH$_3(7_{4,3}-6_{3,4})$AA             & 48  & 347.916 &  C$_2$H$_5$OH$(20_{4,17}-19_{4,16})$         & 251 \\
337.825 & HC$_3$N$(37-36)v_7=1$                        & 629 & 347.983 &  UL                                                \\
337.838 & CH$_3$OH$(20_{6,14}-21_{5,16})$E             & 676 & 348.261 &  CH$_3$CH$_2$CN$(39_{2,37}-38_{2,36})$       & 344 \\
337.878 & CH$_3$OH$(7_{1,6}-6_{1,5})$A($v_t$=2)        & 748 & 348.340 &  HN$^{13}$C(4--3)                            & 42  \\
337.969 & CH$_3$OH$(7_{1,6}-6_{1,5})$A($v_t$=1)        & 390 & 348.345 &  CH$_3$CH$_2$CN$(40_{2,39}-39_{2,38})$       & 351 \\
338.081 & H$_2$CS$(10_{1,10}-9_{1,9})$                 & 102 & 348.532 &  H$_2$CS$(10_{1,9}-9_{1,8})$                 & 105 \\
338.125 & CH$_3$OH$(7_{0,7}-6_{0,6})$E                 & 78  & 348.910 &  HCOOCH$_3(28_{9,20}-27_{9,19})$E            & 295 \\
338.143 & CH$_3$CH$_2$CN$(37_{3,34}-36_{3,33})$        & 317 & 348.911 & CH$_3$CN$(19_{9}-18_{9})$                    & 745 \\
338.306 & SO$_2$$(14_{4,14}-18_{3,15})$                & 197 & 349.025 & CH$_3$CN$(19_{8}-18_{8})$                    & 624 \\
338.345 & CH$_3$OH$(7_{1,7}-6_{1,6})$E                 & 71  & 349.107 & CH$_3$OH$(14_{1,13}-14_{0,14})$              & 43  \\
338.405 & CH$_3$OH$(7_{6,2}-6_{6,1})$E                 & 244 \\
\hline
\hline
\multicolumn{6}{l}{\footnotesize$^a$ The detection of this CH$_3$OH $v_t=2$ line is doubtful since other close $v_t=2$ lines with similar}\\
\multicolumn{6}{l}{\footnotesize excitation temperatures were not detected.}
\label{linelist}
\end{longtable}

\end{document}